\documentclass{ws-ijmpd}

\usepackage{lmodern}

\usepackage[T1]{fontenc}
\usepackage[latin9]{inputenc}
\usepackage{units}
\usepackage{amsmath}
\usepackage{amssymb}
\usepackage{graphicx}
\usepackage{esint}
\usepackage[super,compress]{cite}

\begin{document}

\markboth{Paulo Luz, Jos\'e P. S. Lemos}{Newtonian wormholes
with spherical symmetry and tidal forces on test particles}

\title{Newtonian wormholes with spherical symmetry and tidal forces on
  test particles}

\author{Paulo Luz}
\address{Centro Multidisciplinar de Astrof\'{i}sica - CENTRA,
Departamento de F\'{i}sica, Instituto Superior T\'{e}cnico - IST,
Universidade de Lisboa - UL, Av. Rovisco Pais 1, 1049-001
Lisboa,Portugal \\Email: paulo.luz@ist.utl.pt}
\author{Jos\'{e} P. S. Lemos}
\address{Centro Multidisciplinar de Astrof\'{i}sica - CENTRA,
Departamento de F\'{i}sica, Instituto Superior T\'{e}cnico - IST,
Universidade de Lisboa - UL, Av. Rovisco Pais 1, 1049-001
Lisboa,Portugal \\Email: joselemos@ist.utl.pt}


\maketitle

\begin{abstract}
A spherically symmetric wormhole in Newtonian gravitation in curved
space, enhanced with a connection between the mass density and the
Ricci scalar, is presented. The wormhole, consisting of two connected
asymptotically flat regions, inhabits a spherically symmetric curved
space. The gravitational potential, gravitational field and the pressure
that supports the fluid that permeates the Newtonian wormhole are
computed. Particle dynamics and tidal effects in this geometry are
studied. The possibility of having Newtonian black holes in this theory
is sketched. 
\end{abstract}



\section{Introduction}

The idea that Newton's theory of gravitation, can be formulated in
curved space has been recently analyzed by Abramowicz and collaborators
\cite{Abramowicz1,Abramowiczellis}$\,$. In addition to considering Newtonian
gravitation in curved space, an equation linking the geometry of the
3-space with the matter sector was put forward \cite{ehlers,ellis}$\,$.
This is called an enhanced Newtonian gravitation. In particular, this
modification to Newton's theory allows the construction of a wormhole
space \cite{Lemos_Luz}$\,$. Wormholes in general relativity have been
studied in several works see 
\cite{morristhorne,visserbook,lemosetal,balakinlemos,lemosdias}
for example. In this paper, we report on the construction of a spherically
symmetric wormhole \cite{Lemos_Luz} in this enhanced Newtonian gravitation.
We further analyze the tidal effects that emerge from the gravitational
forces and the curvature of space in this spherically symmetric Newtonian
wormhole space.

The outline of the paper is as follows: in Sec.~\ref{fund} we start
by writing the equations that define the enhanced Newtonian theory
of gravitation. In Sec.~\ref{newtonworm} we construct a static spherically
symmetric Newtonian wormhole, we find the gravitational field, the
gravitational potential and the pressure of the fluid that supports
the wormhole. Then, we study test particle's motion in the wormhole
geometry and gravitational field. Finally we analyze the tidal forces
exerted by the gravitational field and the curvature of space in two
nearby particles. In Sec.~\ref{conc}, we conclude and speculate
on the possible existence of truly Newtonian black holes in this enhanced
Newtonian theory of gravitation.

\section{The fundamental equations\label{fund}}

The classical formulation of Newton's theory of gravitation was constructed
for an absolute 3-dimensional Euclidean space. We see, however, that
the set of equations that comprise Newton's theory of gravitation
are also well defined for curved space. Indeed, Abramowicz et al.
\cite{Abramowicz1,Abramowiczellis} recently proposed a formulation
of Newton's theory of gravitation in curved space.

Poisson's equation in curved space is given by 
\begin{equation}
g^{ij}\nabla_{i}\nabla_{j}\phi=4\pi\, G\,\rho\,,\label{curvedpois}
\end{equation}
where $g_{ij}$ is the curved space metric, $\nabla_{i}$ is the covariant
derivative induced by the
metric, the indices $i,j$ run as $i,j=1,2,3$, and $\rho$ is the
density of the matter. For static systems the continuity equation
is trivially verified and the Euler equation is simplified to 
\begin{equation}
\nabla_{i}\, p+\rho\nabla_{i}\,\phi=0\,,\label{eulereq}
\end{equation}
where $p$ is the fluid's pressure. Now, a possible enhancement to
Newton's theory of gravitation was proposed in \cite{Abramowiczellis}
where a relation between the geometry of space and the matter was
introduced and given by 
\begin{equation}
R=2k\rho\,,\label{connectgeomwithmatter}
\end{equation}
where $R$ is the Ricci scalar and $k$ is an arbitrary constant.
The equations of motion of a test particle with mass $m$ subjected
to a gravitational potential $\phi$ are given by Newton's second
law, 
\begin{equation}
m\, a^{i}=-m\, g^{ij}\nabla_{j}\phi\,,\label{curvedeom}
\end{equation}
where it was assumed that the inertial and gravitational mass of the
test particle are equal. Eqs.~\eqref{curvedpois}-\eqref{curvedeom}
define an enhanced Newtonian gravitation.

\section{Spherical Newtonian wormholes \label{newtonworm}}

\subsection{Construction of the spherical Newtonian wormhole \label{sub:Construction}}

\subsubsection{Matter density and metric}

Our purpose is to make use of Eqs.~\eqref{curvedpois}-\eqref{connectgeomwithmatter}
to construct a static, spherical symmetric wormhole space. Let us
then generically write the metric of the space as 
\begin{equation}
ds^{2}=A\left(r\right)dr^{2}+r^{2}\left(d\theta^{2}+\sin^{2}\theta\, d\varphi^{2}\right)\,,
\end{equation}
such that the Ricci scalar $R$ is 
\begin{equation}
R=\frac{2\left[\left(A\left(r\right)-1\right)A\left(r\right)+rA'\left(r\right)\right]}{r^{2}A\left(r\right){}^{2}}\,.\label{eq:5}
\end{equation}
Now, to proceed we consider the expression for the mass density to
be 
\begin{equation}
\rho\left(r\right)=\alpha e^{-\frac{r^{2}}{b^{2}}}\left(2-\frac{b^{2}}{r^{2}}\right)\,,\label{eq:mass_density}
\end{equation}
where $\alpha$ has dimensions of mass density and $b$ of distance.
Substituting Eqs.~\eqref{eq:5} and \eqref{eq:mass_density} in Eq.~\eqref{connectgeomwithmatter}
we find the general form of the metric, for the imposed mass density,
\begin{equation}
ds^{2}=\frac{r}{r+\beta\, re^{-\frac{r^{2}}{b^{2}}}+C_{1}}\, dr^{2}+r^{2}\left(d\theta^{2}+\sin^{2}\theta\, d\varphi^{2}\right)\,,\label{eq:semi_general_metric}
\end{equation}
where $C_{1}$ is an integration constant and 
\begin{equation}
\beta\equiv b^{2}k\alpha\,.\label{eq:beta}
\end{equation}

\subsubsection{The embedding: the value of the 
integration constant and the parameter
$\beta$}

Having 
found the generic form of the metric, we have now to restrict the
values of the parameters $C_{1}$, $\alpha$ and $k$ to have a wormhole
geometry. For such, we will follow~\cite{morristhorne} and
use an embedding diagram. So, in the Euclidean embedding space the
axially symmetric embedded surface can be, using cylindrical coordinates
$(\bar{r},z,\bar{\varphi})$, uniquely described by a function $z(r)$.
Identifying the coordinates $\left(\bar{r},\bar{\varphi}\right)$
of the embedding Euclidean space with the coordinates $\left(r,\varphi\right)$
of the wormhole space, we find the following relation 
\begin{equation}
\frac{dz}{dr}=\pm\sqrt{\frac{r}{r+\beta\, 
re^{-\frac{r^{2}}{b^{2}}}+C_{1}}-1}\,,
\label{eq:embedding_equation}
\end{equation}
which can be used to study the properties of the space. 
From Eq.~\eqref{eq:embedding_equation}
and imposing the throat condition we find that the integration constant
$C_{1}$ is given by 
\begin{equation}
C_{1}=-b\left(1+\frac{\beta}{e}\right)\,,\label{eq:C_1}
\end{equation}
and the range of values of the parameter $\beta\equiv k\alpha\, b^{2}$
is constrained to be $-\infty<\beta<e$. The same restriction to the
parameter $\beta$ is found from the flare-out condition. Substituting
Eq.~\eqref{eq:C_1} in Eq.~\eqref{eq:semi_general_metric} we find
\begin{equation}
ds^{2}=\frac{1}{1-\frac{b(r)}{r}}dr^{2}+
r^{2}\left(d\theta^{2}+\sin^{2}\theta d\varphi^{2}\right)\,,
\label{eq:final_metric_r}
\end{equation}
where 
\begin{equation}
b\left(r\right)=b+\frac{\beta}{e}\left(b-re^{1-
\frac{r^{2}}{b^{2}}}\right)\,.\label{eq:shape_function}
\end{equation}
Another restriction to have a wormhole geometry is that although
near the wormhole's throat, $r=b$, the radial coordinate $r$ is
ill behaved, the proper radial distance must be finite. This condition
implies a new restriction to $\beta$
\begin{equation}
-\infty<\beta\leqslant\beta_{{\rm crit}}\,,\label{eq:beta_range}
\end{equation}
where the critical value of $\beta$ 
is given by
$\beta_{{\rm crit}}\equiv\inf_{1\,\leq\,\frac{r}{b}\,
<\,\infty}\left[\frac{\frac{r}{b}-1}{e^{-1}-\frac{r}{b}\, 
e^{-\frac{r^{2}}{b^{2}}}}\right]$,
and $\inf$ represents the infimum of the function in the specified
range. This gives 
\begin{equation}
\beta_{\rm crit}=2.338\,,\label{eq:beta_crit}
\end{equation}
up to the third decimal place.
Comparing Eq.~\eqref{eq:beta_crit} with the restriction found from the
throat condition, $\beta<e$, we conclude that the product
$\beta=k\alpha\, b^{2}$ is then restricted by
Eqs.~\eqref{eq:beta_range} and \eqref{eq:beta_crit}.

Due to the range of values that the parameter $\beta$ might take,
various distinct cases could be considered. Here we shall be interested
in the class of wormholes whose parameters $\alpha$, $k$ and\textbf{
}$b$ obey $0<\beta\leqslant\beta_{{\rm crit}}$ together with $\alpha>0$,
and so $k>0$.

\subsubsection{Removal of the coordinate singularity}

\noindent Now that we have found the metric of the Newtonian wormhole
we have to remove the coordinate singularity at $r=b$. Considering
a new coordinate $l$, defined as $l^{2}=r^{2}-b^{2}$,
such that
$-\infty<l<\infty$,
one finds that 
the metric of the space can be rewritten as
\begin{equation}
ds^{2}=\frac{e\, l^{2}}{\left(b^{2}+l^{2}\right)\left(e+\beta\,
e^{-\frac{l^{2}}{b^{2}}}\right)-b\sqrt{b^{2}+l^{2}}
\left(e+\beta\right)}dl{}^{2}+\left(l^{2}+b^{2}\right)
\left(d\theta^{2}+\sin^{2}\theta\,
d\varphi^{2}\right)\,.\label{eq:final_metric_l}
\end{equation}
It will also be useful to write the mass density in the new coordinates.
It is given by the expression 
\begin{equation}
\rho\left(l\right)=\alpha\frac{e^{-1-\frac{l^{2}}{b^{2}}}
\left(b^{2}+2l^{2}\right)}{b^{2}+l^{2}}\,.\label{eq:densitl_l}
\end{equation}

\subsection{Gravitational field, gravitational potential and pressure
support of the spherical Newtonian wormhole
\label{sub:Gravitational_field}}

\begin{figure}
\centering\includegraphics[scale=0.5]{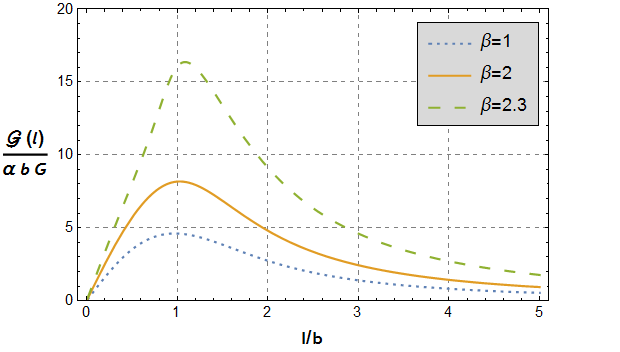} 
\protect
\caption{\label{fig:1}
Magnitude of the gravitational force field for various
values of the parameter $\beta\equiv b^{2}k\alpha$ as a function
of the radial coordinate $l$ (here assume $l\geq0$).
When $\beta=\beta_{\rm crit}=2.338$ the 
magnitude of the gravitational force field tends to infinity
at some value of $l/b$.
}
\end{figure}

Having defined the geometry of the space we can now study the
gravitational potential of the Newtonian wormhole by solving
Eq.~\eqref{curvedpois}.  It is, however, considering the symmetries of
the system, straightforward to solve Eq.~\eqref{curvedpois} by using
Gauss's law for gravity and integrating this equation over a volume
whose boundary are the surfaces of constant gravitational
potential. Defining the gravitational force field as
\begin{equation}
\mathcal{G}^{i}\equiv g^{ij}\,\phi_{,\, j}\,,\label{gravforce}
\end{equation}
with a comma denoting a partial derivative, we find 
\begin{equation}
\mathcal{G}^{i}\left(l\right)=\frac{G\, 
m\left(l\right)}{2\left(b^{2}+l^{2}\right)}
\sqrt{\frac{\left(b^{2}+l^{2}\right)\left(
e+\beta\, e^{-\frac{l^{2}}{b^{2}}}\right)-
b\sqrt{b^{2}+l^{2}}\left(e+\beta\right)}{e\, 
l^{2}}}\,\delta_{l}^{i}\,,
\label{eq:Grav_field}
\end{equation}
where the mass within radius $l$ is given by 
\begin{equation}
m\left(l\right)=4\pi\alpha b^{3}\int_{1}^{1
+\nicefrac{l^{2}}{b^{2}}}\frac{e^{-x}\left(2
x-1\right)}{\sqrt{x\left(1+\beta e^{-x}\right)
-\left(1+\frac{\beta}{e}\right)\sqrt{x}}}\, dx\,.
\label{eq:masswithinl_integral}
\end{equation}
Defining the magnitude of the gravitational field as 
\begin{equation}
\mathcal{G}=\sqrt{g_{ij}\,\mathcal{G}^{i}\,\mathcal{G}^{j}}\,,
\label{mag2}
\end{equation}
we give in Fig.~\ref{fig:1} its behavior as a function of the radial
coordinate $l$ for various values of the parameter $\beta$. 

Using the expression for the gravitational field,
Eq.~\eqref{eq:Grav_field} together with Eq.~\eqref{gravforce}, the
gravitational potential $\phi$ can be computed at a given point using
numerical methods, provided the wormhole parameters are given.
Moreover, these equations can also be used to find the pressure of the
fluid that covers the wormhole space. Numerically solving
Eqs.~\eqref{eulereq} and \eqref{eq:Grav_field} we find that the
pressure is positive throughout the space showing that the wormhole is
hold against gravitational collapse by pressure \cite{Lemos_Luz}$\,$.
From Eq.~(\ref{gravforce}) one sees that the gradient of the potential
$\phi$ is the gravitational force field $\mathcal{G}^{i}$, and from
Eq.~(\ref{eulereq}) the gradient of the pressure $p$ goes with minus
the gradient of the gravitational potential $\phi$. Thus one can infer
from Fig.~\ref{fig:1}, and the help of Eqs.~(\ref{eulereq}),
(\ref{gravforce}), and (\ref{mag2}), that at the wormhole's center
$l=0$, i.e., at the throat, the pressure is maximal, as one would
expect.

In the limiting case that the wormhole
has $\beta=\beta_{{\rm crit}}$ a Newtonian event horizon develops at
some radial coordinate $l_{h}>0$, in the sense that any particle
inside the sphere defined by $l_{h}>0$ can only affect an outside
observer if it has infinite acceleration, and thus infinite
velocity. Thus, 
since no particles can come out of this inner region one is
in the presence of a Newtonian black hole. However, for
$\beta=\beta_{{\rm crit}}$ the pressure support at the throat goes to
infinity, and so, in this limit, the wormhole eventually
collapses.

\subsection{Test particle dynamics and motion in 
the spherical wormhole gravitational
field\label{sub:Test_particle_dynamics}}

\subsubsection{The equations of motion}

Now we use Eq.~\eqref{curvedeom} to study the motion of a test particle
in the gravitational field of the wormhole space. Consider that a
particle's path is described by a curve $\gamma$, whose components
$x^{i}$ are given by $x^{i}=\left(l\left(t\right),\theta\left(t\right),
\varphi\left(t\right)\right)$
such that the particle's velocity is $v^{i}\equiv\dot{x}^{i}$,
where a dot means a derivative with respect to time $t$.
The
right hand side of Eq.~\eqref{curvedeom} is the gravitational field
given in Eq.~\eqref{eq:Grav_field} and the left hand side the acceleration
of the particle, defined as $a^{i}=v^{j}\,\nabla_{j}\, v^{i}$. 
It is possible to show that one can treat the 
problem of the motion of a test particle by considering 
pure equatorial 
orbits, i.e., $\theta=\frac\pi2$, $\dot\theta=0$ and
$\ddot\theta=0$. 
Gathering
these results the equations of motion are 
\begin{equation}
\begin{aligned}\ddot{l}+\left[\frac{b\, e^{\frac{l^{2}}{b^{2}}}\left(2b^{2}+l^{2}\right)\left(e+\beta\right)-2\sqrt{b^{2}+l^{2}}\left[\beta\, l^{2}\left(1+l^{2}/b^{2}\right)+b^{2}\left(e^{1+\frac{l^{2}}{b^{2}}}+\beta\right)\right]}{2l\left(b^{2}+l^{2}\right)\left[b\, e^{\frac{l^{2}}{b^{2}}}\left(e+\beta\right)-\sqrt{b^{2}+l^{2}}\left(e^{1+\frac{l^{2}}{b^{2}}}+\beta\right)\right]}\right]\,\,\dot{l}{}^{2}+\\
\hspace{70bp}+\left[\frac{b\sqrt{b^{2}+l^{2}}\left(e+\beta\right)-\left(b^{2}+l^{2}\right)\left(e+\beta\, e^{-\frac{l^{2}}{b^{2}}}\right)}{e\, l}\right]\,\,\dot{\varphi}{}^{2}=\\
\hspace{70bp}=-\frac{G\, m\left(l\right)}{2\sqrt{e}\,\left(b^{2}+l^{2}\right)l}\sqrt{\left(b^{2}+l^{2}\right)\left(e+\beta\, e^{-\frac{l^{2}}{b^{2}}}\right)-b\sqrt{b^{2}+l^{2}}\left(e+\beta\right)}\,,
\end{aligned}
\label{eq:ldotdot_simp}
\end{equation}
\begin{equation}
\theta=\frac{\pi}{2}\,,\quad\dot{\theta}=0\,,\label{eq:thetadotdot_simp}
\end{equation}
\begin{equation}
\ddot{\varphi}+2\,\frac{l}{b^{2}+l^{2}}\,\,\dot{l}\,\dot{\varphi}=0\,.\label{eq:phidotdot_simp}
\end{equation}

\subsubsection{Solutions of the equations of motion}

Having found the equations of motion of a test particle, we have now to solve
them. Let us consider the simpler problem of a test particle describing
pure circular motion in the Newtonian wormhole space. In this case
the coordinate $l$ is a constant, $l_{0}$, say, so $\dot{l}=0$
and $\ddot{l}=0$. From this, Eq.~\eqref{eq:ldotdot_simp} can be
solved for $\dot{\varphi}$ and obtain 
\begin{equation}
\dot{\varphi}=\sqrt{\frac{\sqrt{e}\, G\, m\left(l_{0}\right)}{2\left(b^{2}+l_{0}^{2}\right)\sqrt{\left(b^{2}+l_{0}^{2}\right)\left(e+\beta\, e^{-\frac{l_{0}^{2}}{b^{2}}}\right)-b\sqrt{b^{2}+l_{0}^{2}}\left(e+\beta\right)}}}\,.\label{eq:circular_phi}
\end{equation}
Eq.~\eqref{eq:circular_phi} relates the radial position $l_{0}$
and the angular velocity $\dot{\varphi}$ for a particle to describe
a circular orbit in the wormhole space. Now, except the case of pure
circular motion, to solve the equations of motion \eqref{eq:ldotdot_simp}-\eqref{eq:phidotdot_simp}
for a more generic motion, numerical methods must be used, provided
the initial position and velocity of the test particle and the wormhole
parameters \textbf{$b$ }and $\alpha$ are given.

\subsection{Tidal effects from the gravitational field and the geometry of the
spherical Newtonian wormhole}

\subsubsection{Tidal 
deviation equation for the spherical Newtonian wormhole}

We now analyze the relative separation of two test particles in the
gravitational field of the spherical Newtonian wormhole space. The two
test particles are initially considered to be infinitesimally close,
subjected only to the gravitational field of the Newtonian wormhole,
such that the acceleration of a test particle at a certain point is
given by Eq.~\eqref{curvedeom} and
Eqs.~\eqref{eq:Grav_field}-\eqref{eq:masswithinl_integral}.  This
relative separation, i.e., the tidal deviation equation, is in part
given by the variation of the gravitational force field acted on each
particle and another part given by the geodesic deviation due to the
spatial curvature through the equation,
\begin{equation}
\frac{D^{2}n^{i}}{dt^{2}}=n^{k}\left(-
\nabla_{k}\mathcal{G}^{i}+R_{jlk}^{i}v^{j}v^{l}\right)\,,
\label{eq:deviation_equation}
\end{equation}
where $n^{i}$ is the separation vector between two infinitesimally
close particles, $\mathcal{G}^{i}$ is the gravitational force field
generated by the wormhole mass, given by Eq.~\eqref{eq:Grav_field},
$v^{i}$ is the velocity of the fiducial test particle and $R_{jlk}^{i}$
represents the Riemann tensor.

Let us simplify the calculations, by remarking that, as was mentioned
in Sec.~(\ref{sub:Test_particle_dynamics}), the velocity of the
fiducial test particle can  be assumed to be along the equator,
$\theta=\pi/2$ (see Eq.~\eqref{eq:thetadotdot_simp}). Hence, in
Eq.~\eqref{eq:deviation_equation}, we might take all the terms in
$v^{\theta}$ to zero. Notice, however, that this does not mean that,
in general, the acceleration of the component $n^{\theta}$ of the
separation vector is zero since the motion of the second particle
may not be along the equator.
Now, considering the metric of the spherical Newtonian wormhole space,
Eq.~\eqref{eq:final_metric_l}, and assuming $\theta=\pi/2$ and
$\dot{\theta}=0$, we find from Eq.~\eqref{eq:deviation_equation}
that the deviation equations for two infinitesimally close test particles
in the spherically symmetric Newtonian wormhole space are 
\begin{eqnarray}
\frac{D^{2}n^{l}}{dt^{2}}= & -n^{l}\left(\partial_{l}\mathcal{G}^{l}+\Gamma_{ll}^{l}\mathcal{G}^{l}\right)+R_{\varphi l\varphi}^{l}v^{\varphi}\left(n^{\varphi}v^{l}-n^{l}v^{\varphi}\right)\,,\label{dev_l}\\
\frac{D^{2}n^{\theta}}{dt^{2}}= & n^{\theta}\left[-\Gamma_{\theta l}^{\theta}\mathcal{G}^{l}+R_{ll\theta}^{\theta}\left(v^{l}\right)^{2}+R_{\varphi\varphi\theta}^{\theta}\left(v^{\varphi}\right)^{2}\right]\,,\label{dev_theta}\\
\frac{D^{2}n^{\varphi}}{dt^{2}}= & -\Gamma_{\varphi l}^{\varphi}\mathcal{G}^{l}n^{\varphi}+R_{ll\varphi}^{\varphi}v^{l}\left[n^{\varphi}v^{l}-n^{l}v^{\varphi}\right]\,.\label{eq:dev_phi}
\end{eqnarray}
where the Christoffel symbols that appear in Eqs.~\eqref{dev_l}-\eqref{eq:dev_phi}
have the form 
\begin{equation}
\Gamma_{ll}^{l}=\frac{b^{3}(\beta+e)e^{\frac{l^{2}}{b^{2}}}\left(2b^{2}+l^{2}\right)-2\sqrt{b^{2}+l^{2}}\left[l^{2}\beta\left(b^{2}+l^{2}\right)+b^{4}\left(e^{\frac{l^{2}}{b^{2}}+1}+\beta\right)\right]}{2b^{2}l\left(b^{2}+l^{2}\right)\left[b(\beta+e)e^{\frac{l^{2}}{b^{2}}}-\sqrt{b^{2}+l^{2}}\left(e^{\frac{l^{2}}{b^{2}}+1}+\beta\right)\right]}\,,\label{Gammalll}
\end{equation}
\begin{equation}
\Gamma_{\theta l}^{\theta}=\Gamma_{\varphi l}^{\varphi}=\frac{l}{b^{2}+l^{2}}\,.\label{Gammathetaltheta}
\end{equation}
The other non-zero Christoffel symbols necessary to calculate the
Riemann tensor components are $\Gamma_{\varphi\varphi}^{l}=\Gamma_{\theta\theta}^{l}\sin^{2}\theta$,
$\Gamma_{\varphi\theta}^{\varphi}=\cot\theta$, and
$\Gamma_{\varphi\varphi}^{\theta}=-\Gamma_{\varphi\theta}^{\varphi}\sin^{2}\theta$.
At the plane $\theta=\pi/2$ one has 
\begin{equation}
\Gamma_{\theta\theta}^{l}=\Gamma_{\varphi\varphi}^{l}=-\frac{\left(b^{2}+l^{2}\right)\left(\beta e^{-\frac{l^{2}}{b^{2}}}+e\right)-b(\beta+e)\sqrt{b^{2}+l^{2}}}{el}\,,
\end{equation}
and $\Gamma_{\varphi\theta}^{\varphi}=\Gamma_{\varphi\varphi}^{\theta}=0$,
the remaining Christoffel symbols being zero or given by symmetry
of the ones just calculated. The Riemann tensor components appearing
in Eqs.~\eqref{dev_l}-\eqref{eq:dev_phi}, assuming $\theta=\pi/2$,
are 
\begin{eqnarray}
R_{\varphi l\varphi}^{l}= & \frac{2\beta e^{-\frac{l^{2}}{b^{2}}}\sqrt{\left(b^{2}+l^{2}\right)^{3}}-b^{3}(\beta+e)}{2eb^{2}\sqrt{b^{2}+l^{2}}}\,,\\
R_{ll\theta}^{\theta}=R_{ll\varphi}^{\varphi}= & \frac{l^{2}\left(2\beta\sqrt{\left(b^{2}+l^{2}\right)^{3}}-b^{3}(\beta+e)e^{\frac{l^{2}}{b^{2}}}\right)}{2b^{2}\left(b^{2}+l^{2}\right)^{2}\left(b(\beta+e)e^{\frac{l^{2}}{b^{2}}}-\sqrt{b^{2}+l^{2}}\left(e^{\frac{l^{2}}{b^{2}}+1}+\beta\right)\right)}\,,\\
R_{\varphi\varphi\theta}^{\theta}= & \frac{\beta e^{-\frac{l^{2}}{b^{2}}}\sqrt{b^{2}+l^{2}}-b(\beta+e)}{e\sqrt{b^{2}+l^{2}}}\,.\label{eq:deviation_non_null_Riemann}
\end{eqnarray}
These equation complete the geodesic deviation equations \eqref{dev_l}-\eqref{eq:dev_phi}.

\subsubsection{Tidal effects for pure radial motion }

Now, looking at Eqs.~\eqref{dev_l}-\eqref{eq:dev_phi}, in general,
it is not possible to find analytical solutions since the fiducial
test particle's velocity must also be taken into account. So, to solve
this system of differential equations, we must first solve the equations
of motion, Eqs.~\eqref{eq:ldotdot_simp}-\eqref{eq:phidotdot_simp},
and in general there is no analytical solution. We can however work
out some features of the tidal effects in the case of pure radial
motion of two test particles such that throughout the motion the line
that connects the two particles is purely radial, i.e., the case
where $n^{\theta}$, $n^{\varphi}$, $v^{\theta}$, and $v^{\varphi}$
are zero throughout the particles' motion. In this case Eq.~\eqref{dev_l}
simplifies to 
\begin{equation}
\frac{D^{2}n^{l}}{dt^{2}}=-n^{l}\left(\partial_{l}\mathcal{G}^{l}+\Gamma_{ll}^{l}\mathcal{G}^{l}\right)\,,\label{eq:Radial_deviation_l}
\end{equation}
and Eqs.~\eqref{dev_theta} and \eqref{eq:dev_phi} yield $\frac{D^{2}n^{\theta}}{dt^{2}}=0$
and $\frac{D^{2}n^{\varphi}}{dt^{2}}=0$, respectively. Although it
is still not possible to solve analytically Eq.~\eqref{eq:Radial_deviation_l}
we can instead make a qualitative analysis to infer the general behavior
of two initially close test particles describing pure radial motion.

\begin{figure}
\centering\includegraphics[scale=0.5]{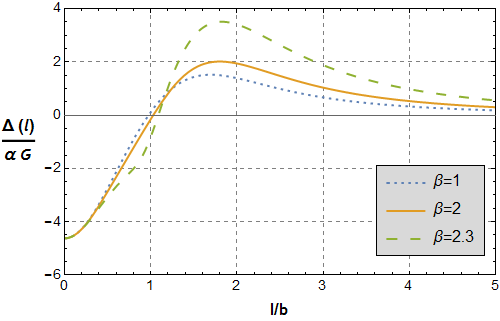}\protect\caption{\label{fig:2}Plot of the quantity $\Delta\equiv-
\partial_{l}\mathcal{G}^{l}-\Gamma_{ll}^{l}\mathcal{G}^{l}$
as a function of the radial coordinate $l$ (here assume $l\geq0$). }
\end{figure}

Defining the right hand side of Eq.~\eqref{eq:Radial_deviation_l}
as 
\begin{equation}
\Delta\left(l\right)\equiv-\partial_{l}\mathcal{G}^{l}-
\Gamma_{ll}^{l}\mathcal{G}^{l}\label{Deltal}
\end{equation}
we present in Fig.~\ref{fig:2} its behavior. We see that the sign of
the function $\Delta$ changes at some value of the coordinate $l$,
$l_{c}$, say. Now, when $l>l_{c}$ the function $\Delta\left(l\right)$
is positive, $\Delta\left(l\right)>0$, which implies that the
separation vector has positive acceleration. So, suppose two test
particles describing pure radial motion, such that the line that
connects them is purely radial. One particle has radial coordinate
$l_{1}$, the other $l_{2}$, and let us assume, $l_{1}<l_{2}$ and
$l_{1,2}>l_{c}$ with $l_{c}>0$ for simplicity. In this case, since
$\Delta\left(l\right)>0$, the particle with coordinate $l_{1}$ is
accelerating more than the particle with coordinate $l_{2}$, and they
fly apart. On the other hand, for $l<l_{c}$ the function
$\Delta\left(l\right)$ is negative, $\Delta\left(l\right)<0$, so the
separation vector has negative acceleration. Assuming now
$l_{1,2}<l_{c}$, and still $l_{1}<l_{2}$, this means that the particle
with coordinate $l_{1}$ is accelerating less, but still towards the
wormhole's throat, than the particle with coordinate
$l_{2}$. Numerically solving Eq.~\eqref{eq:Radial_deviation_l} for two
initially close test particles in the regions $l>l_{c}$ and $l<l_{c}$,
indeed, we verify this conclusions. Notice that as we consider
particles further away from the wormhole's throat we recover the tidal
behavior expected from Newtonian gravitation since the space is
asymptotically flat.

\section{Conclusions \label{conc}}

A static, spherically symmetric wormhole in an enhanced Newtonian
theory of gravitation was constructed. The Newtonian wormhole's mass
density is positive, the gravitational field of the wormhole is
well-behaved and the matter that sustains it has positive
pressure. Test particles' motion in the wormhole gravitational field
and tidal effects were studied.  We have also argued about the
possibility of having true Newtonian black holes, i.e., Newtonian
objects that have regions from which any particle must have infinite
acceleration and thus infinite velocity to escape to the 
outside of it, in this enhanced Newtonian theory of gravitation.

\end{document}